\documentclass[12pt]{article}
\usepackage[letterpaper,top=2cm,bottom=2cm,left=3cm,right=3cm,marginparwidth=1.75cm]{geometry}
\usepackage{xcolor}
\usepackage{amsmath}
\usepackage{graphicx,natbib}
\usepackage[colorlinks=true, allcolors=blue]{hyperref}
\usepackage{multirow}
\usepackage{graphics,booktabs}
\usepackage{amsthm,amssymb,authblk}

\theoremstyle{definition}

\newtheorem{algorithm}{Algorithm}

\usepackage{array}  % for "\extrarowheight" macro

\title{Efficient subsampling for high-dimensional data}
\author[1]{Vasilis Chasiotis}
\author[2]{Lin Wang}
\author[1]{Dimitris Karlis}

\affil[1]{\small{Department of Statistics, Athens University of Economics and Business, Greece}}
\affil[2]{\small{Department of Statistics, Purdue University, USA}}
\date{}

\begin{document}
	
	\maketitle
	
	\begin{abstract}
		In the field of big data analytics, the search for efficient subdata selection methods that enable robust statistical inferences with minimal computational resources is of high importance. A procedure prior to subdata selection could perform variable selection, as only a subset of a large number of variables is active. We propose an approach when both the size of the full dataset and the number of variables are large. This approach firstly identifies the active variables by applying a procedure inspired by random LASSO (Least Absolute Shrinkage and Selection Operator) and then selects subdata based on leverage scores to build a predictive model. Our proposed approach outperforms approaches that already exists in the current literature, including the usage of the full dataset, in both variable selection and prediction, while also exhibiting significant improvements in computing time. Simulation experiments as well as a real data application are provided.
	\end{abstract}
	
	
	\section{Introduction}\label{introduction}
	
	Given the challenges presented by big data due to its volume, variety, and complexity, efficient data analysis methodologies are of high importance. Extracting high-quality information from big data is crucial for meaningful insights and informed decision-making. %Traditional methods often fail due to their computational intensity and inability to scale. This has led to 
	Several innovative techniques have been developed to handle and analyze large datasets with a big sample size. Among these, subsampling methods have emerged as a practical solution, allowing for the analysis of a representative subset of the observations without compromising on accuracy or computational efficiency.
	
	Subsampling approaches, such as the bags of little bootstraps \citep{kleiner2014scalable}, divide-and-conquer strategies \citep{lin2011aggregated,chen2014split,song2015split}, and sequential updating for streaming data \citep{schifano2016online,xue2020online}, address these challenges by breaking down the data into manageable parts. In the divide-and-conquer approach, for instance, statistical analyses are performed on different segments of the data, and the results are synthesized to form comprehensive conclusions. Sequential updating methods, on the other hand, are designed to handle data that is available in streams, enabling real-time analysis without the need to store the entire dataset. These methods significantly reduce the computational burden and make it feasible to apply complex statistical tools to big data.
	
	%Within subsampling methodologies, identifying an optimal subset, or subdata, from the full dataset is critical. The subdata should be representative of the entire dataset, allowing for accurate statistical inference while reducing computational complexity. This is where subdata selection methods play a crucial role. By carefully choosing a small representative sample of size $k$ from the full data size $n$, these methods enable the application of robust statistical analyses with a fraction of the computational resources required for the full dataset. 
	
	Current literature shows a growing interest in optimal subsampling and subdata selection techniques designed to yield precise parameter estimates for specified statistical models trained on a selected subsample, with studied models including linear regression (\citealp{ma2015leveraging,derezinski2018leveraged,wang2019information}; \citealp{wang2021oss,wang2022balanced,chasiotis2024}), logistic regression \citep{wang2018optimal,wang2019more,cheng2020information}, generalized linear models, \citep{fithian2014local,ai2019,yu2020optimal,zhang2021optimal,han2020local}, linear mixed models \citep{zhu2024group}, quasi-likelihood \citep{yu2020optimal}, quantile regression \citep{ai2021optimal,wang2020optimal,fan2021optimal,shao2022optimal}, and nonparametric regression \citep{meng2020more,zhang2024independence}. Also, there are methods for model selection \citep{yu2022} and misspecification \citep{meng2020lowcon}. Additionally, model-free subsampling methods have been proposed \citep{mak2018support,joseph2022split,joseph2021supervised}. A review of optimal subsampling methods can be found in the works of \cite{yao2021review} and \cite{yu2024review}. Optimal subsampling methods significantly reduce the number of observations $n$, leading to faster model training. However, when the number of variables $p$ is also large, these methods can become computationally intensive, limiting their applicability. 
	
	When $n$ and $p$ are both large with \(n\gg p\),  \cite{singh} proposed the efficient method of combining LASSO and subdata selection (CLASS) for variable selection and model training. However, CLASS may conservatively misidentify inactive variables as active, especially as the number of active variables grows. This tendency, exacerbated by correlated data, can lead to slower and less accurate results.
	%addressed the scenario where , but $p$ is still substantial (in the thousands). They assumed that the response can be modeled with a linear model and employed the Information-Based Optimal Subdata Selection (IBOSS) method alongside LASSO to integrate variable selection and subdata selection. The proposed algorithm that provides more details for their method is called CLASS (Combining Lasso and Subdata Selection).
	%While the concept underlying CLASS is promising, it has notable limitations. CLASS does not consistently demonstrate excellent performance across all simulation experiments. Specifically, 
	%However, CLASS may misidentify inactive variables as active, particularly as the number of active variables increases. This issue is exacerbated in the case of correlated data, where the misidentification of inactive variables as active becomes even more pronounced. Additionally, CLASS can be more computationally expensive than SIS-IBOSS in certain situations. Although CLASS is designed for high-dimensional data, its tendency to incorrectly identify inactive variables as active can lead to slower and less accurate results. 
	
	This paper proposes an approach that achieves superior performance while reducing computational time compared to CLASS. 
	%Our proposed method retains CLASS's capability to accurately identify all active variables but improves upon it by ensuring that no inactive variables are identified as active. Furthermore, the computational time of our approach is significantly lower than that of CLASS.
	Our proposed approach is in the same spirit as that of \cite{singh} by first employing a variable selection method inspired by the random LASSO technique to identify the active variables from a large pool. Following this, we utilize leverage scores to select the subdata, which is then used to build a predictive model. The usage of the leverage scores is based on the work of \cite{chasiotis2023levss}, who proposed the deterministic leverage score selection algorithm (LEVSS) by \cite{yu2022} for selecting the most informative data points to estimate unknown parameters in a linear model. The dual-stage process of identifying the active variables and selecting subdata not only enhances the efficiency of variable selection and prediction but also significantly reduces computing time, outperforming existing methods in the literature, including those utilizing the full dataset. Through comprehensive simulation experiments and a real data application, we demonstrate the superior performance of our method, establishing it as a robust and computationally efficient solution for big data analytics.
	
	%The proposed method of \cite{singh} demonstrates superior performance compared to existing methods in the current literature, such as SIS-IBOSS \citep{wang2022efficient}. Therefore, screening variables to identify the active ones, followed by subdata selection using only the identified variables, appears to be a very effective approach. We are motivated by this fact and aim to provide an even more effective approach in the same spirit as \cite{singh}.

	Although our approach is similar in spirit to CLASS, it is distinguished by two novel techniques. The first technique involves variable selection through the application of random LASSO on small subsets of the full dataset via a two-phase LASSO regression, which leverages the strengths of random LASSO over traditional LASSO, addressing several of its limitations. The second innovation is the use of LEVSS \citep{yu2022} for subdata selection, rather than IBOSS \citep{wang2019information} used in CLASS, applied to the identified variables. This approach enhances the determinant of the information matrix for active variables, resulting in subdata that are more informative for estimating unknown parameters.
	
	In Section \ref{framework}, we outline the framework of our study, providing the model that forms the basis of our analysis as well as several existing LASSO estimators. Section \ref{new_approach} introduces the novel methodology proposed in this paper. In Section \ref{simulation} we compare the proposed approach with competing approaches on simulated data. Section \ref{real_data} demonstrates the application of our method to real-world data. Finally, in Section \ref{conclusion}, we summarize our findings and offer concluding remarks. 
	
	\section{The framework}\label{framework}
	\subsection{The model}\label{model}
	Let $(\textbf{x}_i, y_i)$, $i=1,\ldots,n$ be the full dataset of $n$ observations,  where $\textbf{x}_i = (x_{i1}, \ldots, x_{ip})^T$ is a $p$-dimensional vector of variables and $y_i$ is the response variable. We consider the following linear regression model:
	\begin{equation}\label{model1}
		y_i = \beta_0 + \beta_1 x_{i1} + \ldots + \beta_p x_{ip} + \varepsilon_i,
	\end{equation}
	where $\beta_0$ is the intercept, $\beta_1, \ldots, \beta_p$ are the slope parameters, and $\varepsilon_i$ is the independent random error with mean $0$ and variance $\sigma^2$.
	
	The matrix form of the model in \eqref{model1} is given by:
	\begin{equation}\label{model2}
		\textbf{y} = \textbf{Z}\boldsymbol{\beta} + \boldsymbol{\varepsilon},
	\end{equation}
	where $\textbf{y} = (y_1, \ldots, y_n)^T$, $\textbf{X} = (\textbf{x}_1,\ldots,\textbf{x}_p)^T$, $\textbf{Z} = (\textbf{1}, \textbf{X})$, $\textbf{1}$ is a $n\times1$ vector of one's, $\boldsymbol{\beta} = (\beta_0, \beta_1, \ldots, \beta_p)^T$, and $\boldsymbol{\varepsilon} = (\varepsilon_1, \ldots, \varepsilon_n)^T$. The ordinary least squares (OLS) estimator of $\boldsymbol{\beta}$, which is also its best linear unbiased
	estimator is
	\[
	\hat{\boldsymbol{\beta}}=\left(\textbf{Z}^T\textbf{Z}\right)^{-1}\textbf{Z}^T\textbf{y}
	\]
	and the covariance matrix of $\hat{\boldsymbol{\beta}}$ is 
	\[
	\sigma^2\left(\textbf{Z}^T\textbf{Z}\right)^{-1}.
	\]
	
	Considering that \{$s_1$,\ldots,$s_k$\} is a subset of \{$1$,\ldots,$n$\}, then $(\textbf{X}_s, \textbf{y}_s)$ denotes the subdata of size $k$ taking from the full dataset $(\textbf{X}, \textbf{y})$ that contains data points $(\textbf{x}_{s_i}, y_{s_i})$. The OLS estimator based on the subdata is given by 
	\[
	\hat{\boldsymbol{\beta}_s}=\left(\textbf{Z}_s^T\textbf{Z}_s\right)^{-1}\textbf{Z}_s^T\textbf{y}_s,
	\]
	where $\textbf{Z}_s = (\textbf{1}_s, \textbf{X}_s)$ and $\textbf{1}_s$ is a $k\times1$ vector of one's. The covariance matrix of $\hat{\boldsymbol{\beta}_s}$, which is the best linear unbiased estimator of $\boldsymbol{\beta}$ based on the subdata $(\textbf{X}_s, \textbf{y}_s)$, is 
	\[
	\sigma^2\left(\textbf{Z}_s^T\textbf{Z}_s\right)^{-1}.
	\]
	
	\subsection{The LASSO estimators}\label{model}
	In case that the number of $p$ is large, under the assumption of effect sparsity, the underlying relationship between the predictors and the response variable involves only a small subset of predictors having non-zero effects, while the remaining predictors have no effect at all. In the context of linear regression models, this assumption implies that the true coefficients associated with most predictors are zero. 
	
	In such situations, we need to apply penalized regression methods in order to analyze the data. For instance, in the LASSO method, proposed by \cite{tibshirani}, the effect sparsity assumption is enforced by penalizing the absolute values of the coefficients, shrinking many of them to zero. This leads to a sparse model retaining only a subset of predictors with non-zero coefficients. The LASSO criterion penalizes the $L_1$-norm of the regression coefficients, that is the LASSO estimator is a solution to the convex optimization problem:
	\[\mathop{\text{arg min}}_{\boldsymbol{\beta}}\left\{\sum_{i=1}^{n}\left(y_i-\textbf{z}_i^T\boldsymbol{\beta}\right)^2+\lambda\sum_{j=1}^{p}|\beta_j|\right\},\]
	where $\textbf{z}_i=\left(1,\textbf{x}_i^T\right)^T$ and $\lambda$ is a nonnegative tuning parameter.
	
	Cross-validation is commonly employed for tuning the parameter $\lambda$. There are two solutions that are widely used. In one of them, the value of $\lambda$ gives the minimum cross-validation error. In the second solution, which is the most regularized one, the value of $\lambda$ is the largest one within one standard error of the minimum cross-validation error.
	
	The LASSO method has two limitations in practice \citep{zou2005}, even though it has been successfully applied in many situations. The first one is that LASSO tends to select only one or some of the variables, shrinking the remaining to zero, when the variables are highly correlated. The second limitation is that, when $p>n$, at most only $n$ variables can be identified by LASSO before it saturates.
	
	To overcome the two aforementioned limitations of LASSO method, several other methods have been proposed in the literature, including the elastic-net \citep{zou2005}, the adaptive LASSO \cite{zou2006}, the relaxed LASSO \citep{meinshausen2007}, variable inclusion and shrinkage algorithms \citep{radchenko2008} and the random LASSO \citep{wang2011}.
	
	The random LASSO proposed by \cite{wang2011} addresses several shortcomings present in LASSO, elastic-net, and similar methods. This method offers improved handling of highly correlated variables by either excluding them entirely or including them all in the model. Also, it maintains a high degree of flexibility in estimating coefficients, especially allowing for different signs among variables. Moreover, random Lasso selects variables without being limited by the sample size.
	
	\section{The proposed approach}\label{new_approach}
	The goal of our method is to build a predictive model based on a subset of data selected from the full dataset having already identified the active variables. Then we fit a linear regression model with the
	variables that have been identified as active, and we obtain the OLS estimators.
	
	A detailed description of our approach is provided in Algorithm \ref{alg}. First of all, Algorithm \ref{alg} performs variable selection (from Step 1 to Step 7). Multiple random samples are drawn from the full dataset by sampling without replacement. LASSO is implemented on these multiple selected subdata among all variables. The importance of all variables is obtained based on the estimators of their coefficients. Then another set of multiple random samples is drawn from the full dataset by sampling without replacement. Now some of the variables are randomly selected with a selection probability which is based on their importance. LASSO is implemented on the new set of the multiple selected subdata among the selected variables. Like \cite{singh}, we use kmeans clustering to obtain the active variables. In Step 8, LEVSS is used to obtain subdata from the full dataset among the active variables. In Step 9, a linear regression model with the active variables is fitted using the subdata selected in Step 8, and the OLS estimators are obtained.
	
	The primary goal of the variable selection procedure outlined in Steps 1-7 is to ensure that no inactive variables are misidentified as active, as well as to accurately identify all active variables as active. To achieve effective variable selection, we implement the concept of random LASSO on small subsets of the full dataset. We expect that active variables will be identified as active by the proposed variable selection, since the active variable are of higher importance and so they will be prioritized for selection in the second phase of the LASSO regression. On the other hand, applying the two-phase LASSO regression we aim to exclude the inactive variables from the second phase, thereby preventing the misidentification of inactive variables as active. As discussed in Section \ref{introduction}, CLASS is particularly prone to misidentifying inactive variables as active, especially in the presence of correlated data. Our proposed variable selection technique seeks to address this limitation, and we anticipate success because random LASSO can effectively handle highly correlated data.
	
	For the subdata selection procedure among the active variables, we use LEVSS to select subdata that are more informative for estimating the unknown parameters in a linear model compared to the subdata selected by CLASS. LEVSS selects data points based on leverage scores derived from the design matrix. Since a data point’s leverage score reflects its influence on the model fit, points with high leverage scores have a stronger impact on parameter estimation. By using LEVSS, we select data points with the highest leverage scores, which provide more information about the variance in the data, thereby reducing variance in the estimated slope parameters. In contrast, in CLASS algorithm, IBOSS selects data points based on the largest and smallest values of each covariate, without directly prioritizing data points that maximize their influence on slope parameter estimates. While the subdata selected by IBOSS include extreme values, which can be useful for variance estimation in some cases, this approach does not necessarily account for how these data points influence the overall structure of the design matrix in linear regression. As a result, IBOSS may lead to poor coverage of the full covariate space.
	
	\begin{algorithm}\label{alg}
		\textit{Step} 1. Draw $n_1$ random samples with size $s$ by sampling without replacement from the full dataset.
		
		\textit{Step} 2. Apply LASSO to obtain estimators $\hat{\beta}^{(r_1)}_j$ for $\beta_j$, $j = 1,\ldots, p$ for the $r_1$-th random sample, $r_1=1, \ldots, n_1$. The estimators for coefficients of these variables that are excluded by LASSO are zero.
		
		\textit{Step} 3. Compute the importance measure of the $j$-th variable by $m_j = \left| \dfrac{\sum_{r_1=1}^{n_1} \hat{\beta}^{(r_1)}_j}{n_1} \right|$.
		
		\textit{Step} 4. Draw another set of $n_2$ random samples with size $s$ by sampling without replacement from the full dataset.
		
		\textit{Step} 5. Select randomly $p_s$ candidate variables with selection probability of the $j$-th variable proportional to its importance $m_j$ obtained in Step 3, and apply LASSO for the $r_2$-th random sample, $r_2=1, \ldots, n_2$. The variables outside the subset of $p_s$ variables or excluded by LASSO are considered as inactive.
		
		\textit{Step} 6. Save the selected active variables to the set $A_{r_2}$.  
		
		\textit{Step} 7. Use kmeans clustering to form two clusters for the counts in  $C = (C_1, \ldots, C_p)$, where $C_j$ is the number of sets $A_1, \ldots, A_{r_2}$ that contain the $j$-th variable. The variables with counts in the cluster with the largest mean are considered as active variables.
		
		\textit{Step} 8. Utilize LEVSS on the active variables identified in Step 7 to select subdata of size $k$.
		
		\textit{Step} 9. Fit a linear regression model with intercept and the
		active variables identified in Step 7 on the subdata selected in Step 8, and obtain the OLS estimators $\hat{\boldsymbol{\beta}_s}=\left(\hat{\beta}_0^{(s)},\hat{\boldsymbol{\beta}}_1^{(s)}\right)$, where $\hat{\boldsymbol{\beta}}_1^{(s)}=\left(\hat{\beta}_1^{(s)},\ldots,\hat{\beta}_{p_a}^{(s)}\right)$ and $p_a$ is the number of active variables.
	\end{algorithm}
	
	In Steps 2 and 5, in which LASSO is implemented, the regularization parameter $\lambda$ used corresponds to the largest value such that error is within one standard error of the minimum in 10-fold cross-validation.
	
	In Step 3, the importance of variables is computed by the estimators of their coefficients (Step 2). Then, in Step 5, the selection probability of variables is determined by their importance. In Step 5 the number of $p_s$ may be larger than the number of $m_j$'s that are not equal to zero. In this case, all variables for which the corresponding $m_j$'s are not equal to zero are selected in Step 5 as candidate variables.
	
	Also, in Step 9, as shown in \cite{wang2019information} and \cite{wang2021oss}, the intercept is estimated by the adjusted estimator $\hat{\beta}_0=\bar{y}-\bar{\textbf{x}}^{T}\hat{\boldsymbol{\beta}}_1^{(s)}$, where $\bar{y}$ and $\bar{\textbf{x}}$ are the mean of the response variable and the vector of means of active variables in the full dataset, respectively.
	
	In Section \ref{simulation}, we demonstrate that choosing $n_1=50$, $n_2=100$, and $p_s=0.1p$ for Algorithm \ref{alg} yields good results. The effects of tuning these parameters are discussed in the Supplementary Material. The simulation studies in Section \ref{simulation} confirm that Algorithm \ref{alg} successfully selects all active variables only, highlighting its superiority over other competing methods in variable selection and prediction.
	
	\section{Simulation studies}\label{simulation}
	In this section, we use simulated data in order to evaluate the performance of the proposed approach both in terms of variable selection and prediction accuracy. Moreover, we present the results of the CLASS and the full data approaches.
	
	In this simulation experiment, we have $p=500$ variables, from which $p_1=10,25,50$ are considered as true active variables. The coefficients of the true active variables are equal to one and the independent random error terms follow a normal distribution $N(0,9)$. Also, the correlation matrix $\mathbf{\Sigma}$ for the $p=500$ variables is equal to $\textbf{I}_p$ and $0.5\left(\textbf{J}_p+\textbf{I}_p\right)$, where $\textbf{J}_p$ is a $p\times p$ matrix of $1$'s and $\textbf{I}_p$ is the $p\times p$ identity matrix. 
	
	Observations $\textbf{x}_i$'s follow a multivariate normal distribution $N(\textbf{0},\mathbf{\Sigma})$, a multivariate log-normal distribution $LN(\textbf{0},\mathbf{\Sigma})$, a multivariate $t$ distribution $t_2(\textbf{0},\mathbf{\Sigma})$, and a mixture distribution of $N(\textbf{0},\mathbf{\Sigma})$, $LN(\textbf{0},\mathbf{\Sigma})$, $t_2(\textbf{0},\mathbf{\Sigma})$ and $t_3(\textbf{0},\mathbf{\Sigma})$ with equal proportions.
	
	The response variable is generated from the linear regression model in \eqref{model1} with the intercept to be equal to $1$. The sizes $n$ of the full dataset are $10^4$, $2\cdot10^4$, $4\cdot10^4$, $8\cdot10^4$, and $10^5$.
	
	In terms of variable selection, the approaches are evaluated based on their average power and average error. 
	Power refers to the accuracy of identifying active variables as active, while error pertains to mistakenly declaring inactive variables as active. In terms of prediction accuracy, the approaches are evaluated according to mean squared error (MSE) of the test data, that is $
	\text{MSE} = \frac{1}{n_{\text{t}}} \sum_{i=1}^{n_{\text{t}}} \left( \textbf{z}_{i,\text{t}}^T \boldsymbol{\beta} - \textbf{z}_{i,\text{t}}^T \hat{\boldsymbol{\beta}} \right)^2$, where $n_{t}$ is the number of observations in the test data, $\textbf{z}_{i, t}=(1,\textbf{x}_{i, t}^T)^T$ is the $i$th observation in the test data, and $\hat{\boldsymbol{\beta}}$ is the OLS estimator of $\boldsymbol{\beta}$. The parameters of variables that were not identified as active are equal to zero. We set $n_t=1000$ using the same joint variable distribution for the full data and test data.  
	
	The full data as well as the test data are generated $100$ times. Therefore, we report the average power, the average error and the MSE of the approaches considered over the $100$ replications. Also, we report the computing time of the approaches considered over one replication.
	
	For the approach in which we use the full data, the regularization parameter $\lambda$ used corresponds to the largest value such that error is within one standard error of the minimum in 10-fold cross-validation. For CLASS, we use $nsample=1000$, $ntimes=100$ and $k=1000$. Also, we examine the performance of CLASS in case that $k=0.1n$.

	\subsection{Evaluation of performance}\label{eval_perf}
\begin{figure}[!htb] 
		\centering 
		\includegraphics[width=1\textwidth]{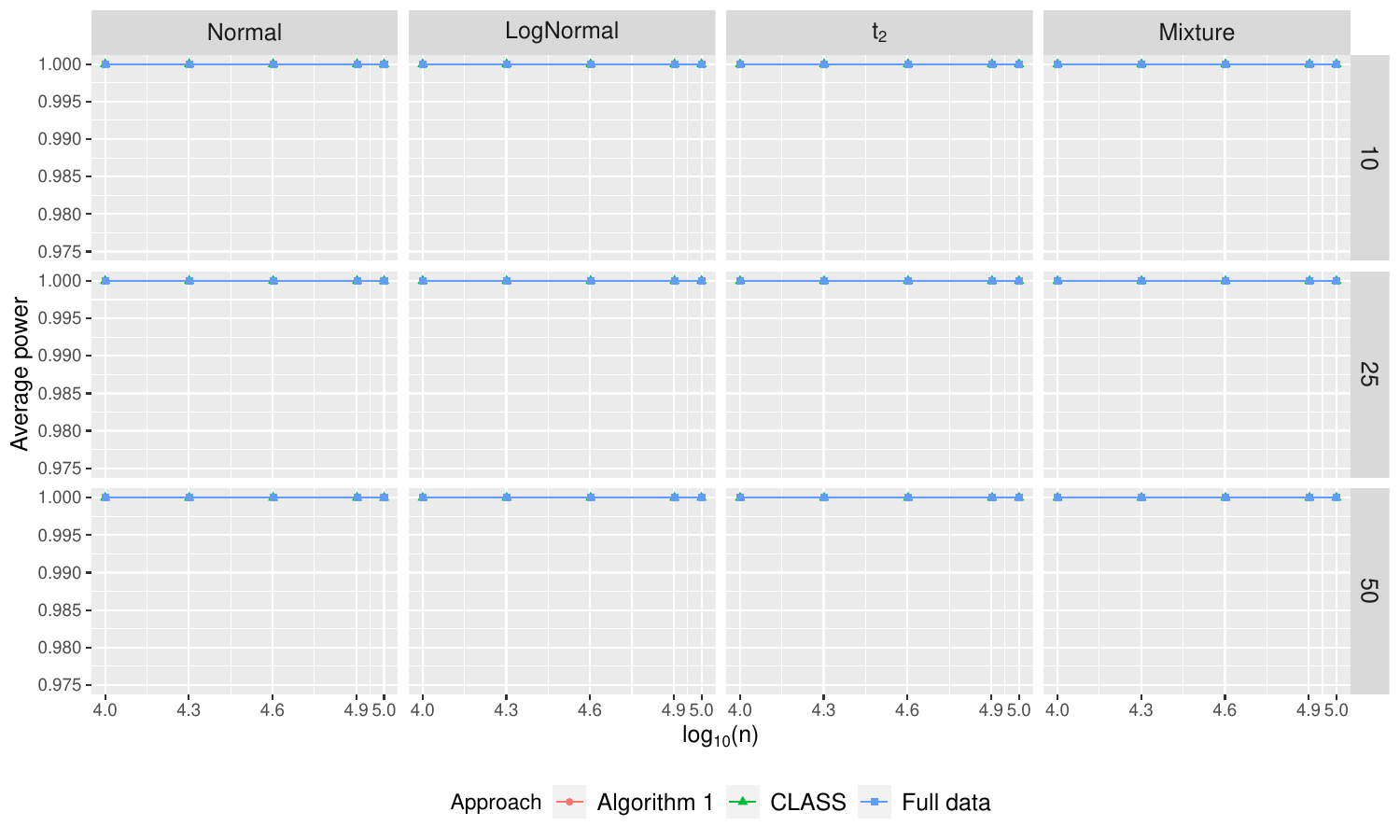} 
		\caption{Average power for $s = 1000$, $\boldsymbol{\Sigma} = \textbf{I}_p $ and $p_1 = 10, 25, 50$.}
		\label{p500sIb1}
	\end{figure}

\begin{figure}[!htb] 
		\centering 
		\includegraphics[width=1\textwidth]{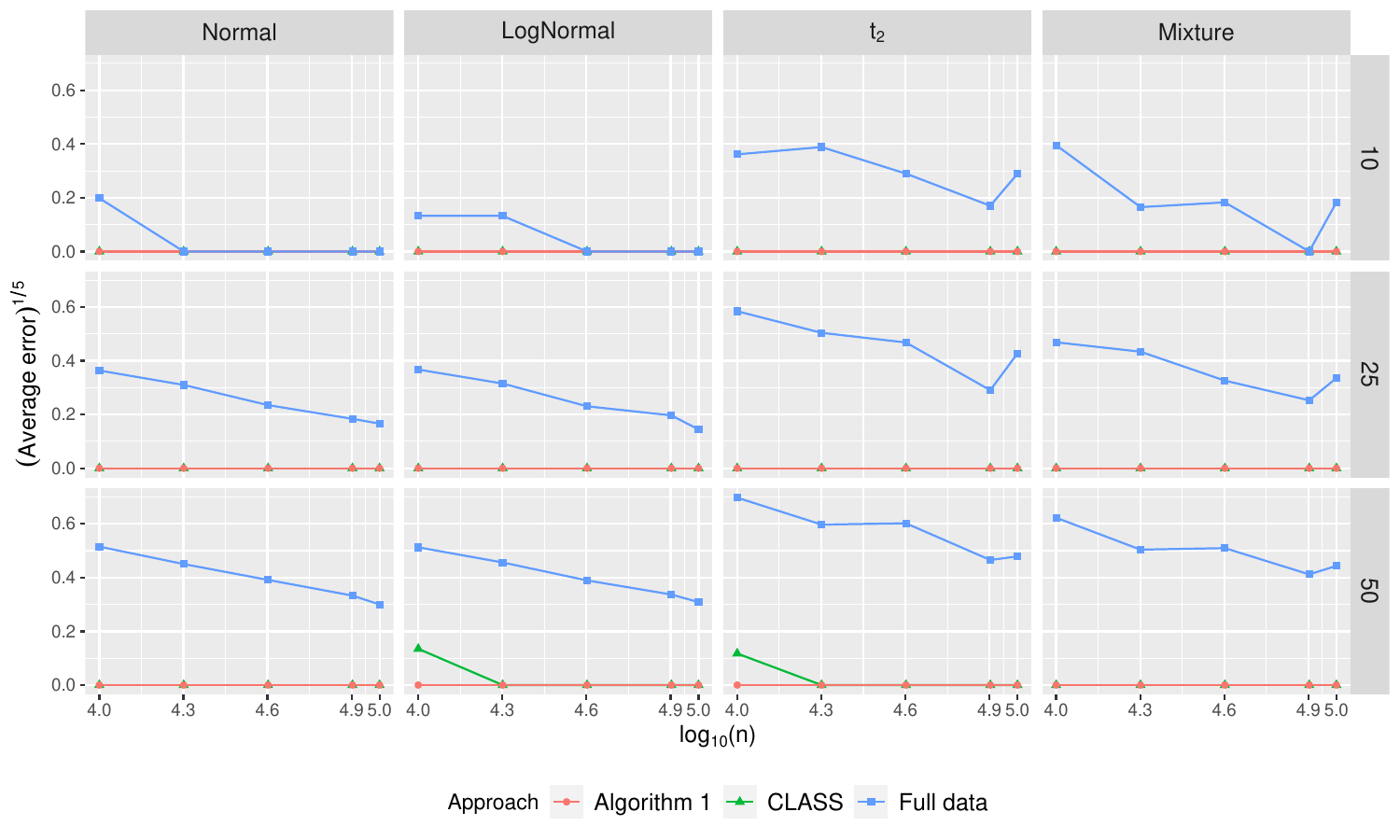} 
		\caption{Average error for $s = 1000$, $\boldsymbol{\Sigma} = \textbf{I}_p $ and $p_1 = 10, 25, 50$.}
		\label{e500sIb1}
	\end{figure}

\begin{figure}[!htb] 
		\centering 
		\includegraphics[width=1\textwidth]{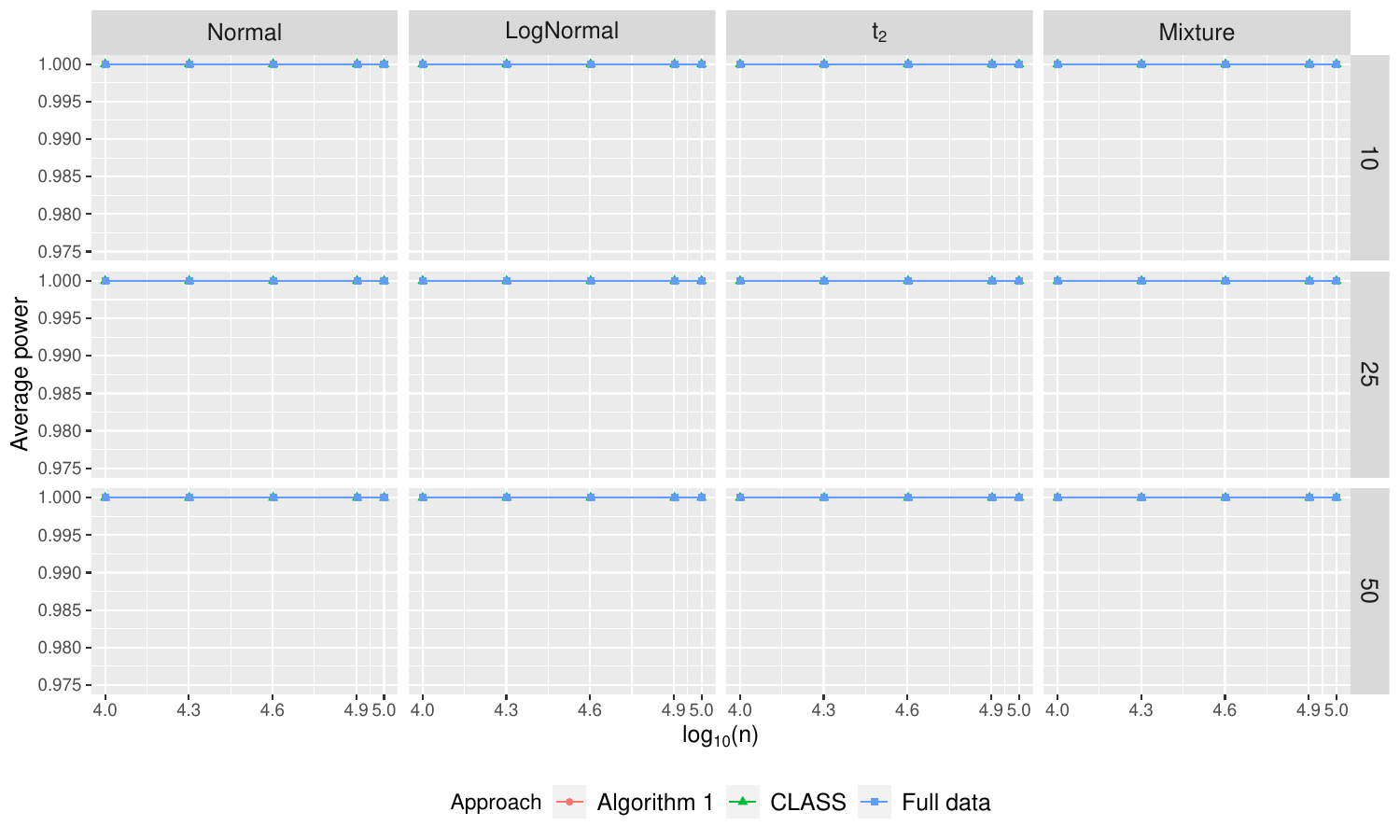} 
		\caption{Average power for $s = 1000$, $\boldsymbol{\Sigma} = 0.5\left(\textbf{J}_p+\textbf{I}_p\right) $ and $p_1 = 10, 25, 50$.}
		\label{p500s05b1}
	\end{figure}

\begin{figure}[!htb] 
		\centering 
		\includegraphics[width=1\textwidth]{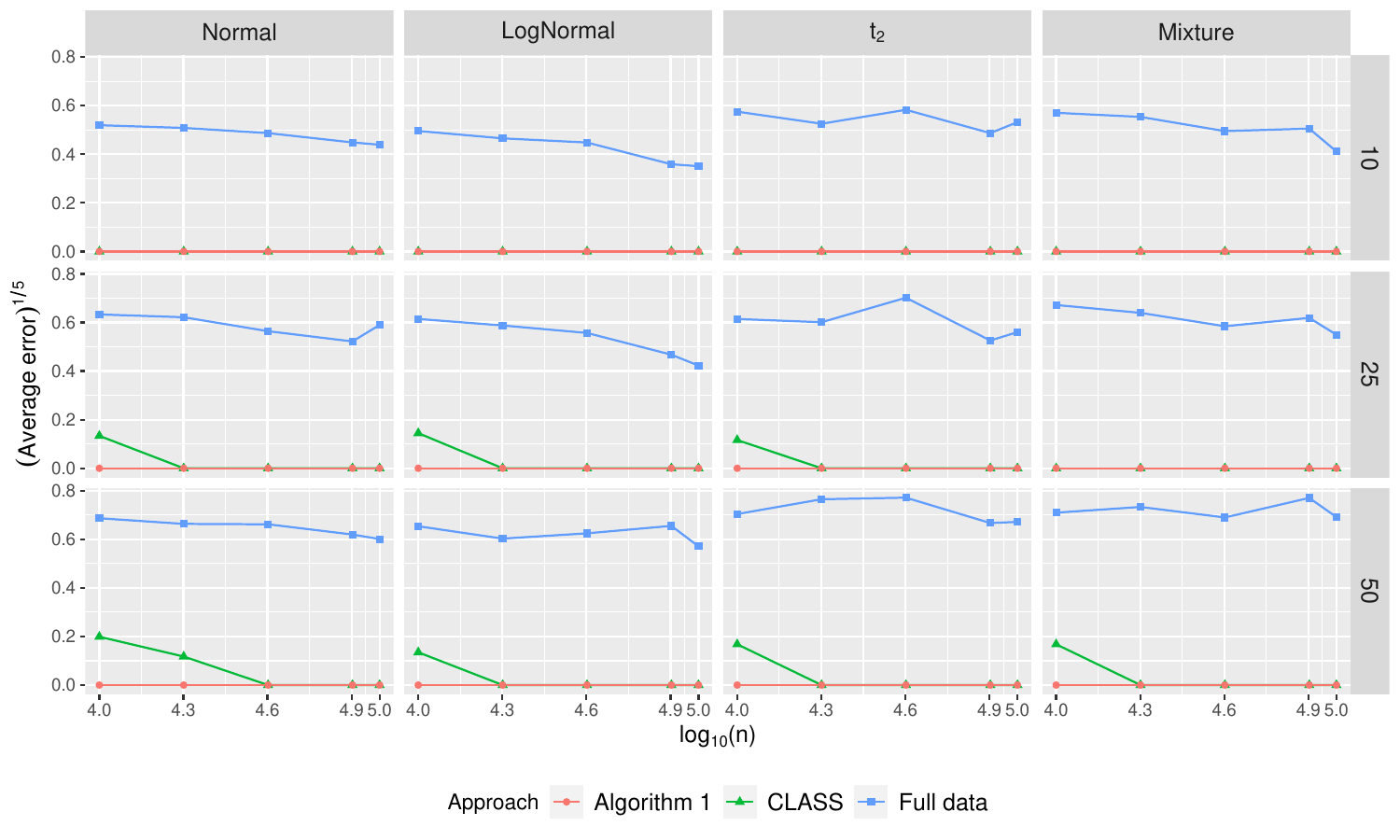} 
		\caption{Average error for $s = 1000$, $\boldsymbol{\Sigma} = 0.5\left(\textbf{J}_p+\textbf{I}_p\right) $ and $p_1 = 10, 25, 50$.}
		\label{e500s05b1}
	\end{figure}

	In Figures \ref{p500sIb1}, \ref{e500sIb1}, \ref{p500s05b1} and \ref{e500s05b1}, the performance of Algorithm \ref{alg} and the approaches of CLASS and the full data for variable selection are presented.
	
	The advantages of Algorithm \ref{alg} compared to the other competing methods are that its power and error are equal to one and zero, respectively, for all distributions considered. CLASS has power close to one for all distributions and a small error in some cases. Also, the full data approach has power close to one but the error is much larger than the one of the Algorithm \ref{alg} and the CLASS, especially when the data are correlated. The results can be considered as expected since Algorithm \ref{alg} selects variables based on a two-phase LASSO regression compared to CLASS that performs only one set of LASSO regressions.
	
\begin{figure}[!htb] 
		\centering 
		\includegraphics[width=1\textwidth]{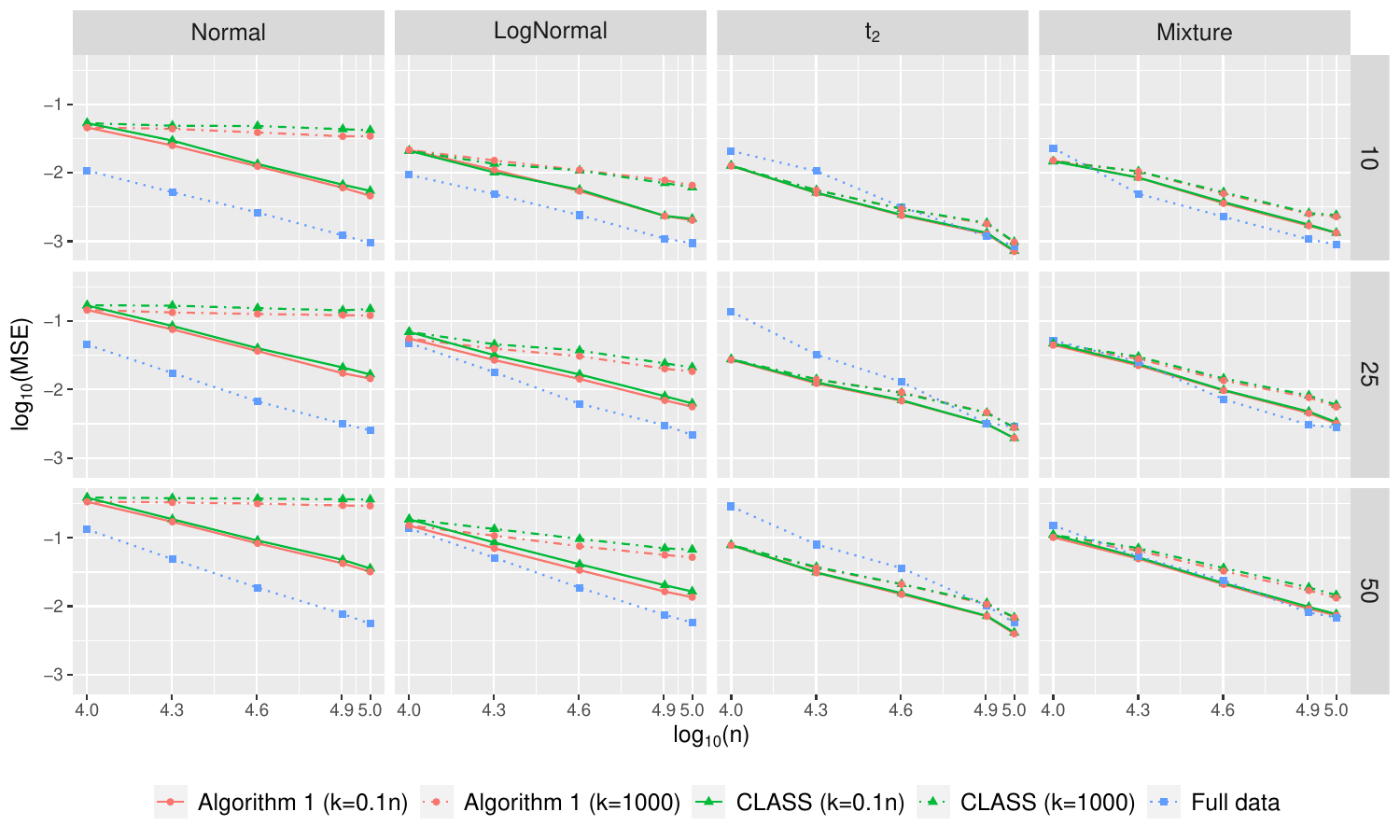} 
		\caption{MSE for $s = 1000$, $\boldsymbol{\Sigma} = \textbf{I}_p $ and $p_1 = 10, 25, 50$.}
		\label{mse500sIb1}
	\end{figure}

\begin{figure}[!htb] 
		\centering 
		\includegraphics[width=1\textwidth]{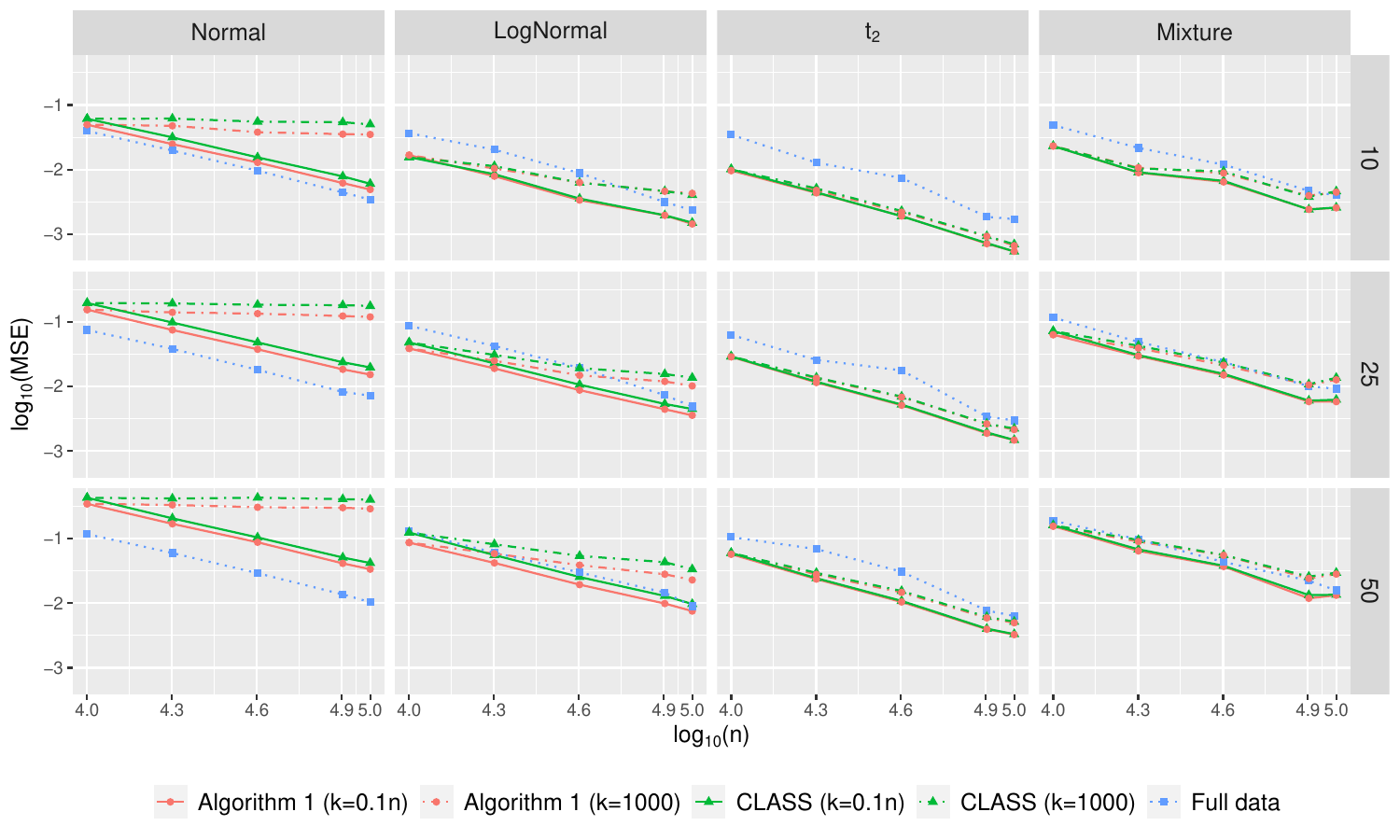} 
		\caption{MSE for $s = 1000$, $\boldsymbol{\Sigma} = 0.5\left(\textbf{J}_p+\textbf{I}_p\right) $ and $p_1 = 10, 25, 50$.} 
		\label{mse500s05b1}
	\end{figure}

	In Figures \ref{mse500sIb1} and \ref{mse500s05b1} the performance of Algorithm \ref{alg} and the approaches of CLASS and the full data for prediction accuracy are presented. 
	
	Another advantage of Algorithm \ref{alg} is that the MSE is the lowest among the considered approaches, for both $k=1000$ and $k=0.1n$. Also, the MSE of Algorithm \ref{alg} and the CLASS approach are close to that of the full data approach, becoming in some cases lower as well, especially for  heavy-tailed distributions. This becomes more intense for both Algorithm \ref{alg} and the CLASS approach when the $p=500$ variables are correlated and the size of the subdata selected based on the variables that have been identified as active in equal to $k=0.1n$. Moreover, the choice of $k=0.1n$ can lead either to lower MSE compared to the one of the full data approach or to a decrease of MSE as the value of $n$ is getting larger.
	
\begin{figure}[!htb] 
		\centering 
		\includegraphics[width=1\textwidth]{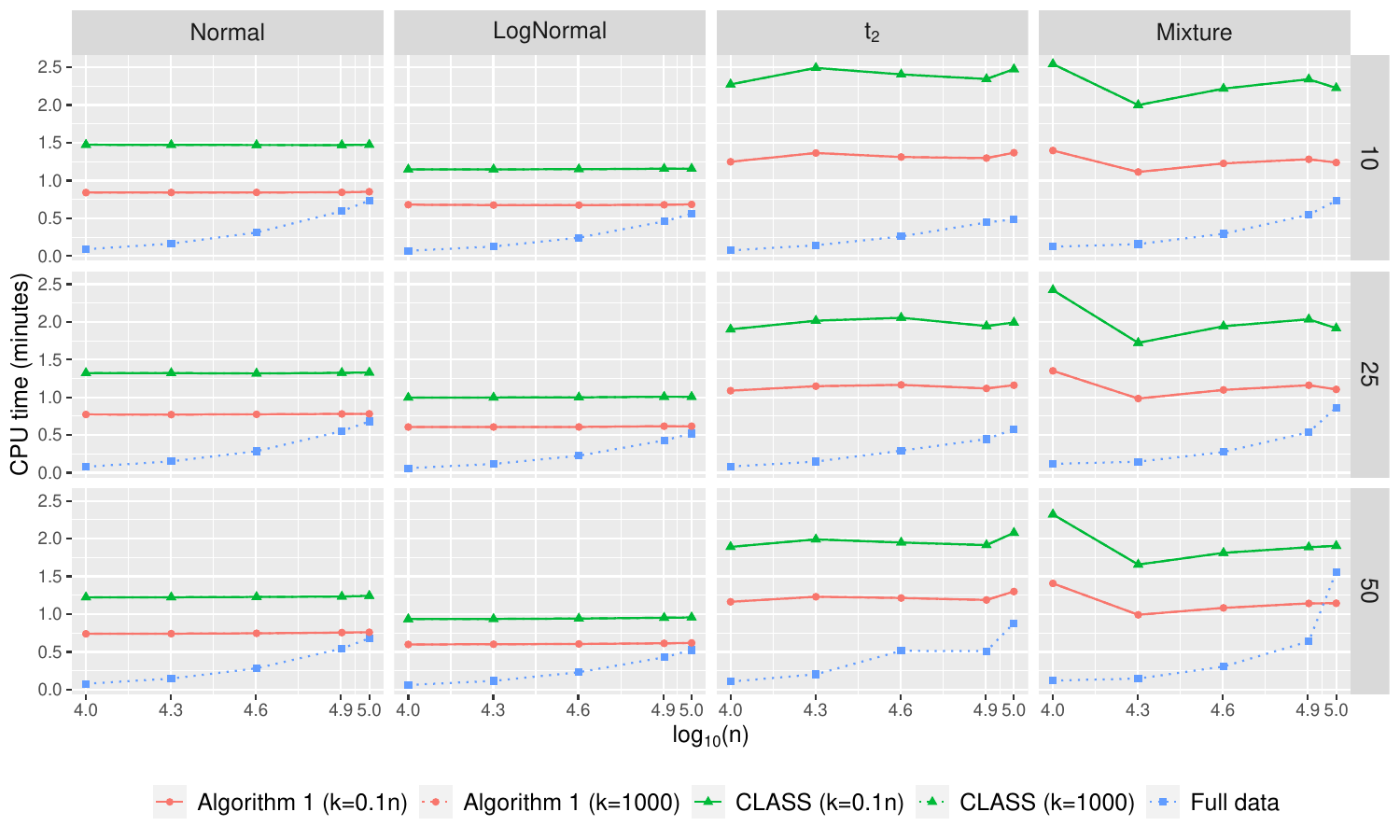} 
		\caption{CPU times (minutes) for $s = 1000$, $\boldsymbol{\Sigma} = \textbf{I}_p $ and $p_1 = 10, 25, 50$.} 
		\label{time500sIb1}
	\end{figure}

\begin{figure}[!htb] 
		\centering 
		\includegraphics[width=1\textwidth]{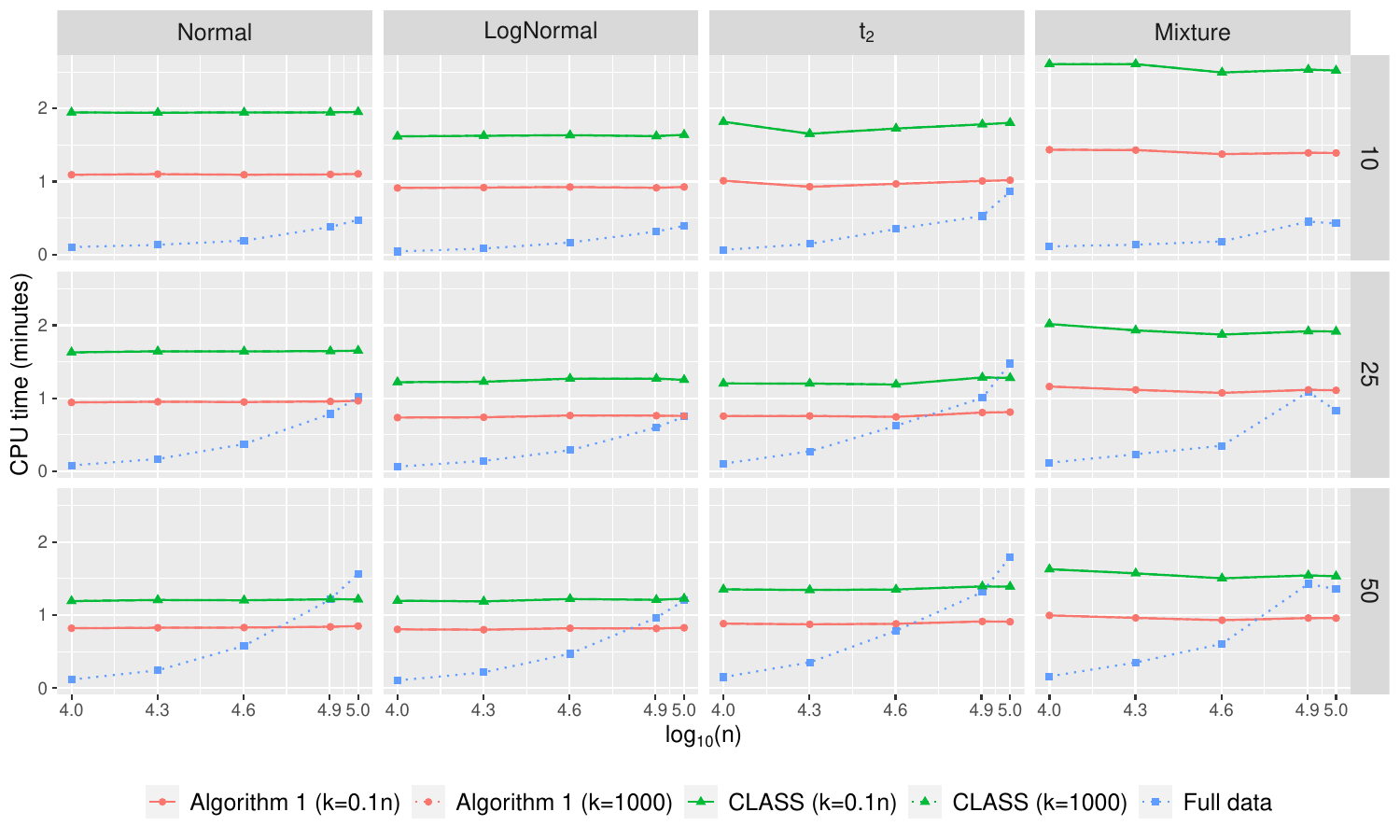} 
		\caption{CPU times (minutes) for $s = 1000$, $\boldsymbol{\Sigma} = 0.5\left(\textbf{J}_p+\textbf{I}_p\right) $ and $p_1 = 10, 25, 50$.} 
		\label{time500s05b1}
	\end{figure}

	In Figures \ref{time500sIb1} and \ref{time500s05b1} the computing times of Algorithm \ref{alg} and the approaches of CLASS and the full data are presented. All computations are carried out on a PC with a 3.6 GHz Intel 8-Core I7 processor and 32GB memory. 
	
	Another advantage of Algorithm \ref{alg} is that it is clearly faster than the CLASS approach and very competitive with full data approach, especially when the $p=500$ variables are correlated, the value of $p_1$ is getting larger, and the value of $n$ is getting larger. The CLASS approach is faster than the full data approach when the values of $p_1$ and $n$ are large. One more advantage of Algorithm \ref{alg} is that it is faster than the full data approach even for smaller values of $p_1$ and $n$. Moreover, we see that the decrease of MSE when $k=0.1n$ instead of $k=100$ for both Algorithm \ref{alg} and the CLASS approach is attained without any additional time.
	
	We note that the MSE of the full data approach is the lowest, particularly for the Normal and Log-normal distributions when the variables are uncorrelated. This also holds true for the Normal distribution when the variables are correlated. However, as the number of $p_1$ increases, the MSE of Algorithm \ref{alg} and CLASS decreases, approaching that of the full data approach for both uncorrelated and correlated variables. Although the full data approach is generally faster than Algorithm \ref{alg} and CLASS, it can become slower as $n$ increases, especially when the variables are correlated. Additionally, it's important to consider that the full data approach may become impractical when $n$ and $p$ are too large, in contrast to Algorithm \ref{alg} and CLASS, which perform variable selection and prediction based on subsets of the full data.
	
	\section{Application to real data}\label{real_data}
	In this real data application, we are dealing with data derived from blog posts, aiming to predict the number of comments a post will receive within the next 24 hours. To mimic real conditions, we establish a baseline time (in the past) and select blog posts published within 72 hours before this baseline. The training dataset consists of $n=52,397$ observations, each comprising $280$ variables representing various attributes of the blog posts at the baseline. These variables encompass details about the source (such as total comments on the source before the baseline and comments in the last 24 hours) and features of the blog post itself (like post length and keyword frequencies). The baseline times for the training data are from 2010 and 2011.
	
	For testing, we have a separate dataset with $7,624$ observations, where the baseline times fall within February and March 2012. This setup mirrors real-world scenarios, where past data are available for training models to predict future events. The provided training and testing splits ensure fair evaluation by maintaining disjoint partitions. Additional information about the dataset can be found at \cite{buza2014} and the ``UCI Machine Learning Repository" \citep{dua2019}.
	
	We are interested in comparing the performance of Algorithm \ref{alg} and the approaches of CLASS and the full data. We consider the mean squared prediction
	error (MSPE) for the testing data for each approach by using $100$ bootstrap samples. Each bootstrap sample is a random sample of size $n$ from the training data using sampling with replacement. For Algorithm \ref{alg}, the predicted values are obtained from the model in Step 9. For CLASS, the predicted values are obtained from the last step of the corresponding algorithm. For the full data approach, to obtain the predicted values, first a LASSO regression model is fitted and then the OLS estimator of the coefficients for the selected variables is obtained. 
	
\begin{figure}[!htb] 
		\centering 
		\includegraphics[width=1\textwidth]{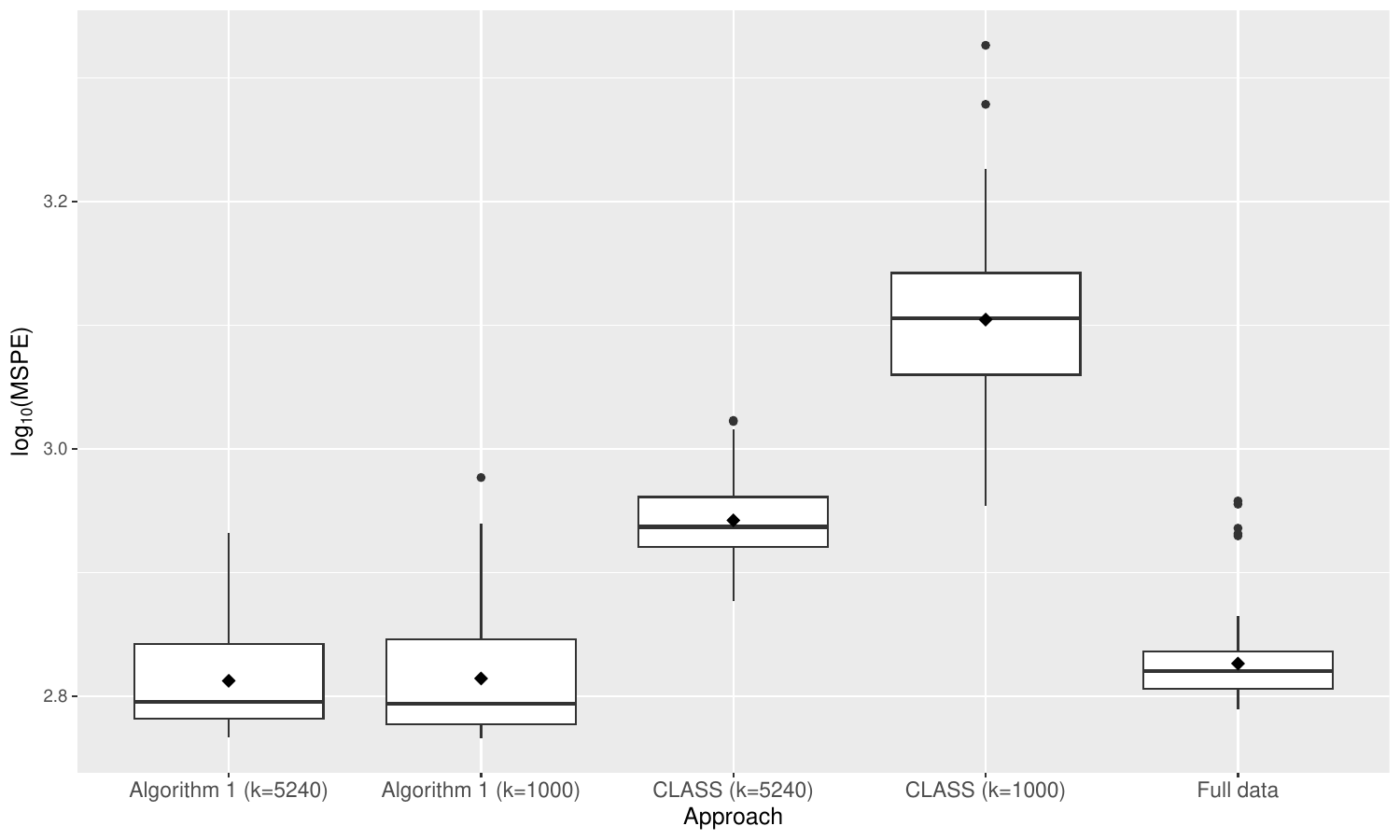} 
		\caption{The bootstrap MSPEs by different approaches over $100$ bootstrap samples on the blog feedback data.} 
		\label{boxplot_real_data}
	\end{figure}

	Figure \ref{boxplot_real_data} shows the bootstrap MSPEs by different approaches. For Algorithm \ref{alg} and the CLASS approach we use $k=1000$ and $k=5240$ (the first integer larger than $0.1n$). The mean values ($\blacklozenge$) are also provided. Algorithm \ref{alg} provides better MSPE than the approaches of CLASS and the full data for both $k=1000$ and $k=5240$. The CLASS approach provides better predictions as the value of $k$ increases from $1000$ to $5240$, but it cannot outperform the approach of the full data. Algorithm \ref{alg} is stable, as the increase in $k$ from $1000$ to $5240$ does not result in significant variation in the prediction.
	
	\section{Concluding remarks}\label{conclusion}
	The subdata selection methods seems to deal very well on identifying the active variables and building a good predictive model, when the number of observations and variables are large. 
	
	In this paper, we proposed a novel approach inspired by \cite{singh} that combines variable selection with subdata selection to enhance the efficiency of predictive modeling, particularly in large-scale datasets. Our method first employs a variable selection strategy inspired by the random LASSO technique to identify active variables from a large pool. We then use leverage scores, as introduced by \cite{chasiotis2023levss}, to select a subset of the most informative data points, which are subsequently used to build a predictive model. This dual-stage process not only improves the efficiency of variable selection and prediction but also significantly reduces computing time, outperforming existing methods, including those utilizing the full dataset.
	
	Our approach addresses several limitations of existing methods, such as CLASS, which does not consistently perform well across all simulations and can be more computationally intensive. By applying the concept of random LASSO on small subsets of the full data, our method overcomes some of the shortcomings of traditional LASSO. Additionally, leveraging LEVSS for subdata selection, as opposed to IBOSS, leads to a larger determinant for the information matrix of the active variables, further enhancing model performance.
	
	The superior statistical performance of our approach, both in variable selection and prediction accuracy, underscores its advantage over competing methods for big data analytics. Our method not only offers faster processing than analyzing the full dataset when the number of observations and variables is large, but it is also applicable in scenarios where full data analysis may be impractical or impossible. This makes our approach a robust and efficient choice for handling complex, large-scale datasets.
	
	\section*{Acknowledgments}
	Lin Wang’s research was supported by NSF grant DMS-2413741 and the Central Indiana Corporate Partnership AnalytiXIN Initiative.

	\bibliographystyle{apalike}
	\bibliography{sample}

\begin{thebibliography}{}

\bibitem[Ai et~al., 2021]{ai2021optimal}
Ai, M., Wang, F., Yu, J., and Zhang, H. (2021).
\newblock Optimal subsampling for large-scale quantile regression.
\newblock {\em Journal of Complexity}, 62:101512.

\bibitem[Ai et~al., 2019]{ai2019}
Ai, M., Yu, J., Zhang, H., and Wang, H. (2019).
\newblock Optimal subsampling algorithms for big data regressions.
\newblock {\em Statistica Sinica}, 31:749--772.

\bibitem[Buza, 2014]{buza2014}
Buza, K. (2014).
\newblock Feedback prediction for blogs.
\newblock {\em Data Analysis, Machine Learning and Knowledge Discovery}, pages
  145--152.

\bibitem[Chasiotis and Karlis, 2023]{chasiotis2023levss}
Chasiotis, V. and Karlis, D. (2023).
\newblock On the selection of optimal subdata for big data regression based on
  leverage scores.
\newblock \url{doi.org/10.48550/arXiv.2305.01597}.

\bibitem[Chasiotis and Karlis, 2024]{chasiotis2024}
Chasiotis, V. and Karlis, D. (2024).
\newblock Subdata selection for big data regression: an improved approach.
\newblock {\em Journal of Data Science, Statistics, and Visualisation}, 4(3).

\bibitem[Chen and Xie, 2014]{chen2014split}
Chen, X. and Xie, M.-G. (2014).
\newblock A split-and-conquer approach for analysis of extraordinarily large
  data.
\newblock {\em Statistica Sinica}, 24(4):1655--1684.

\bibitem[Cheng et~al., 2020]{cheng2020information}
Cheng, Q., Wang, H., and Yang, M. (2020).
\newblock Information-based optimal subdata selection for big data logistic
  regression.
\newblock {\em Journal of Statistical Planning and Inference}, 209:112--122.

\bibitem[Derezinski et~al., 2018]{derezinski2018leveraged}
Derezinski, M., Warmuth, M.~K., and Hsu, D.~J. (2018).
\newblock Leveraged volume sampling for linear regression.
\newblock In {\em Advances in Neural Information Processing Systems},
  volume~31.

\bibitem[Dua and Graff, 2019]{dua2019}
Dua, D. and Graff, C. (2019).
\newblock \uppercase{UCI} machine learning repository.
\newblock \url{http://archive.ics.uci.edu/ml.}

\bibitem[Fan et~al., 2021]{fan2021optimal}
Fan, Y., Liu, Y., and Zhu, L. (2021).
\newblock Optimal subsampling for linear quantile regression models.
\newblock {\em Canadian Journal of Statistics}, 49(4):1039--1057.

\bibitem[Fithian and Hastie, 2014]{fithian2014local}
Fithian, W. and Hastie, T. (2014).
\newblock Local case-control sampling: Efficient subsampling in imbalanced data
  sets.
\newblock {\em Annals of Statistics}, 42(5):1693.

\bibitem[Han et~al., 2020]{han2020local}
Han, L., Tan, K.~M., Yang, T., and Zhang, T. (2020).
\newblock Local uncertainty sampling for large-scale multiclass logistic
  regression.
\newblock {\em The Annals of Statistics}, 48(3):1770--1788.

\bibitem[Joseph and Mak, 2021]{joseph2021supervised}
Joseph, V.~R. and Mak, S. (2021).
\newblock Supervised compression of big data.
\newblock {\em Statistical Analysis and Data Mining: The ASA Data Science
  Journal}, 14(3):217--229.

\bibitem[Joseph and Vakayil, 2022]{joseph2022split}
Joseph, V.~R. and Vakayil, A. (2022).
\newblock Split: An optimal method for data splitting.
\newblock {\em Technometrics}, 64(2):166--176.

\bibitem[Kleiner et~al., 2014]{kleiner2014scalable}
Kleiner, A., Talwalkar, A., Sarkar, P., and Jordan, M.~I. (2014).
\newblock A scalable bootstrap for massive data.
\newblock {\em Journal of the Royal Statistical Society: Series B (Statistical
  Methodology)}, 76(4):795--816.

\bibitem[Lin and Xi, 2011]{lin2011aggregated}
Lin, N. and Xi, R. (2011).
\newblock Aggregated estimating equation estimation.
\newblock {\em Statistics and Its Interface}, 4(1):73--83.

\bibitem[Ma and Sun, 2015]{ma2015leveraging}
Ma, P. and Sun, X. (2015).
\newblock Leveraging for big data regression.
\newblock {\em Wiley Interdisciplinary Reviews: Computational Statistics},
  7(1):70--76.

\bibitem[Mak and Joseph, 2018]{mak2018support}
Mak, S. and Joseph, V.~R. (2018).
\newblock Support points.
\newblock {\em The Annals of Statistics}, 46(6A):2562--2592.

\bibitem[Meinshausen, 2007]{meinshausen2007}
Meinshausen, N. (2007).
\newblock Relaxed lasso.
\newblock {\em Computational Statistics \& Data Analysis}, 52(1):374--393.

\bibitem[Meng et~al., 2020a]{meng2020lowcon}
Meng, C., Xie, R., Mandal, A., Zhang, X., Zhong, W., and Ma, P. (2020a).
\newblock Lowcon: A design-based subsampling approach in a misspecified linear
  model.
\newblock {\em Journal of Computational and Graphical Statistics},
  30(3):694--708.

\bibitem[Meng et~al., 2020b]{meng2020more}
Meng, C., Zhang, X., Zhang, J., Zhong, W., and Ma, P. (2020b).
\newblock More efficient approximation of smoothing splines via space-filling
  basis selection.
\newblock {\em Biometrika}, 107(3):723--735.

\bibitem[Radchenko and James, 2008]{radchenko2008}
Radchenko, P. and James, M.~G. (2008).
\newblock Variable inclusion and shrinkage algorithms.
\newblock {\em Journal of the American Statistical Association},
  103(483):1304--1315.

\bibitem[Schifano et~al., 2016]{schifano2016online}
Schifano, E.~D., Wu, J., Wang, C., Yan, J., and Chen, M.-H. (2016).
\newblock Online updating of statistical inference in the big data setting.
\newblock {\em Technometrics}, 58(3):393--403.

\bibitem[Shao et~al., 2022]{shao2022optimal}
Shao, L., Song, S., and Zhou, Y. (2022).
\newblock Optimal subsampling for large-sample quantile regression with massive
  data.
\newblock {\em Canadian Journal of Statistics}, 51(2):420--443.

\bibitem[Singh and Stufken, 2023]{singh}
Singh, R. and Stufken, J. (2023).
\newblock Subdata selection with a large number of variables.
\newblock {\em The New England Journal of Statistics in Data Science},
  1(3):426--438.

\bibitem[Song and Liang, 2015]{song2015split}
Song, Q. and Liang, F. (2015).
\newblock A split-and-merge bayesian variable selection approach for ultrahigh
  dimensional regression.
\newblock {\em Journal of the Royal Statistical Society: Series B (Statistical
  Methodology)}, 77(5):947--972.

\bibitem[Tibshirani, 1996]{tibshirani}
Tibshirani, R. (1996).
\newblock Regression shrinkage and selection via the lasso.
\newblock {\em Journal of the Royal Statistical Society: Series B
  (Methodological)}, 58:267--288.

\bibitem[Wang, 2019]{wang2019more}
Wang, H. (2019).
\newblock More efficient estimation for logistic regression with optimal
  subsamples.
\newblock {\em Journal of Machine Learning Research}, 20(132):1--59.

\bibitem[Wang and Ma, 2020]{wang2020optimal}
Wang, H. and Ma, Y. (2020).
\newblock Optimal subsampling for quantile regression in big data.
\newblock {\em Biometrika}, 108(1):99--112.

\bibitem[Wang et~al., 2019]{wang2019information}
Wang, H., Yang, M., and Stufken, J. (2019).
\newblock Information-based optimal subdata selection for big data linear
  regression.
\newblock {\em Journal of the American Statistical Association},
  114(525):393--405.

\bibitem[Wang et~al., 2018]{wang2018optimal}
Wang, H., Zhu, R., and Ma, P. (2018).
\newblock Optimal subsampling for large sample logistic regression.
\newblock {\em Journal of the American Statistical Association},
  113(522):829--844.

\bibitem[Wang, 2022]{wang2022balanced}
Wang, L. (2022).
\newblock Balanced subsampling for big data with categorical covariates.
\newblock {\em arXiv preprint arXiv:2212.12595}.

\bibitem[Wang et~al., 2021]{wang2021oss}
Wang, L., Elmstedt, J., Wong, W.~K., and Xu, H. (2021).
\newblock Orthogonal subsampling for big data linear regression.
\newblock {\em Annals of Applied Statistics}, 15(3):1273--1290.

\bibitem[Wang et~al., 2011]{wang2011}
Wang, S., Nan, B., Rosset, S., and Zhu, J. (2011).
\newblock {Random lasso}.
\newblock {\em The Annals of Applied Statistics}, 5(1):468 -- 485.

\bibitem[Xue et~al., 2020]{xue2020online}
Xue, Y., Wang, H., Yan, J., and Schifano, E.~D. (2020).
\newblock An online updating approach for testing the proportional hazards
  assumption with streams of survival data.
\newblock {\em Biometrics}, 76(1):171--182.

\bibitem[Yao and Wang, 2021]{yao2021review}
Yao, Y. and Wang, H. (2021).
\newblock A review on optimal subsampling methods for massive datasets.
\newblock {\em Journal of Data Science}, 19(1):151--172.

\bibitem[Yu et~al., 2024]{yu2024review}
Yu, J., Ai, M., and Ye, Z. (2024).
\newblock A review on design inspired subsampling for big data.
\newblock {\em Statistical Papers}, 65(2):467--510.

\bibitem[Yu and Wang, 2022]{yu2022}
Yu, J. and Wang, H. (2022).
\newblock Subdata selection algorithm for linear model discrimination.
\newblock {\em Statistical Papers}, 63:1883--1906.

\bibitem[Yu et~al., 2020]{yu2020optimal}
Yu, J., Wang, H., Ai, M., and Zhang, H. (2020).
\newblock Optimal distributed subsampling for maximum quasi-likelihood
  estimators with massive data.
\newblock {\em Journal of the American Statistical Association},
  115(530):1--12.

\bibitem[Zhang et~al., 2021]{zhang2021optimal}
Zhang, T., Ning, Y., and Ruppert, D. (2021).
\newblock Optimal sampling for generalized linear models under measurement
  constraints.
\newblock {\em Journal of Computational and Graphical Statistics},
  30(1):106--114.

\bibitem[Zhang et~al., 2024]{zhang2024independence}
Zhang, Y., Wang, L., Zhang, X., and Wang, H. (2024).
\newblock Independence-encouraging subsampling for nonparametric additive
  models.
\newblock {\em Journal of Computational and Graphical Statistics}, pages 1--10.

\bibitem[Zhu et~al., 2024]{zhu2024group}
Zhu, J., Wang, L., and Sun, F. (2024).
\newblock Group-orthogonal subsampling for hierarchical data based on linear
  mixed models.
\newblock {\em Journal of Computational and Graphical Statistics}, pages 1--10.

\bibitem[Zou, 2006]{zou2006}
Zou, H. (2006).
\newblock The adaptive lasso and its oracle properties.
\newblock {\em Journal of the American Statistical Association},
  101(476):1418--1429.

\bibitem[Zou and Hastie, 2005]{zou2005}
Zou, H. and Hastie, T. (2005).
\newblock {Regularization and variable selection via the elastic net}.
\newblock {\em Journal of the Royal Statistical Society Series B: Statistical
  Methodology}, 67(2):301--320.

\end{thebibliography}

\pagebreak
\begin{center}
\textbf{\Large Supplementary Material}
\end{center}

\setcounter{figure}{0}
\setcounter{table}{0}
\setcounter{section}{0}
\makeatletter
\renewcommand{\thetable}{S\arabic{table}}
\renewcommand{\thefigure}{S\arabic{figure}}
\renewcommand{\thesection}{S\arabic{section}}

\section{Tuning parameters of Algorithm 1}
In the Supplementary Material, we present the results for the effect of changing the tuning parameters in the variable selection part of Algorithm 1. 

\begin{figure}[!htb] 
	\centering 
	\includegraphics[width=1\textwidth]{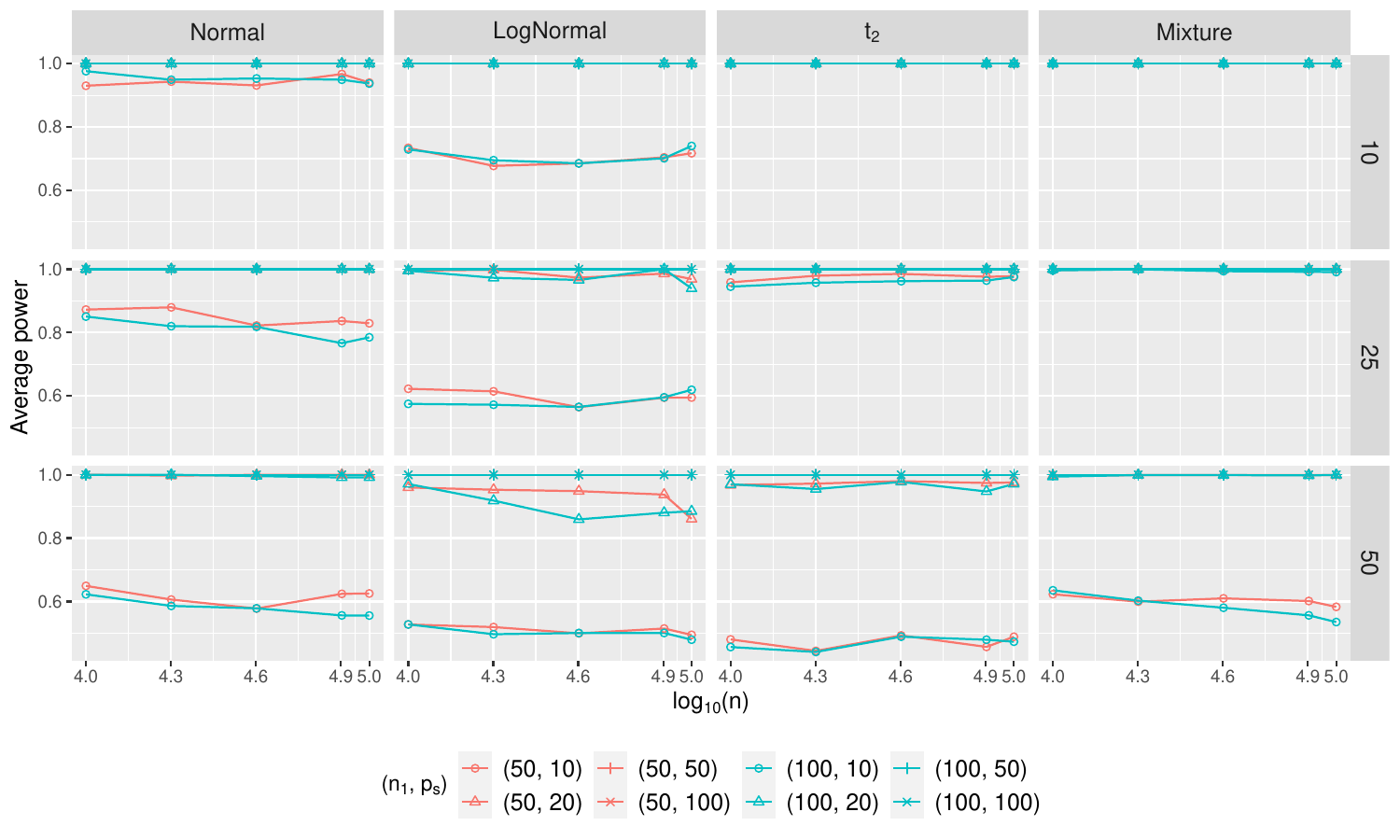} 
	\caption{Average power for $s = 1000$, $\boldsymbol{\Sigma} = \textbf{I}_p $ and $p_1 = 10, 25, 50$.}
	\label{powerI}
\end{figure}

\begin{figure}[!htb] 
	\centering 
	\includegraphics[width=1\textwidth]{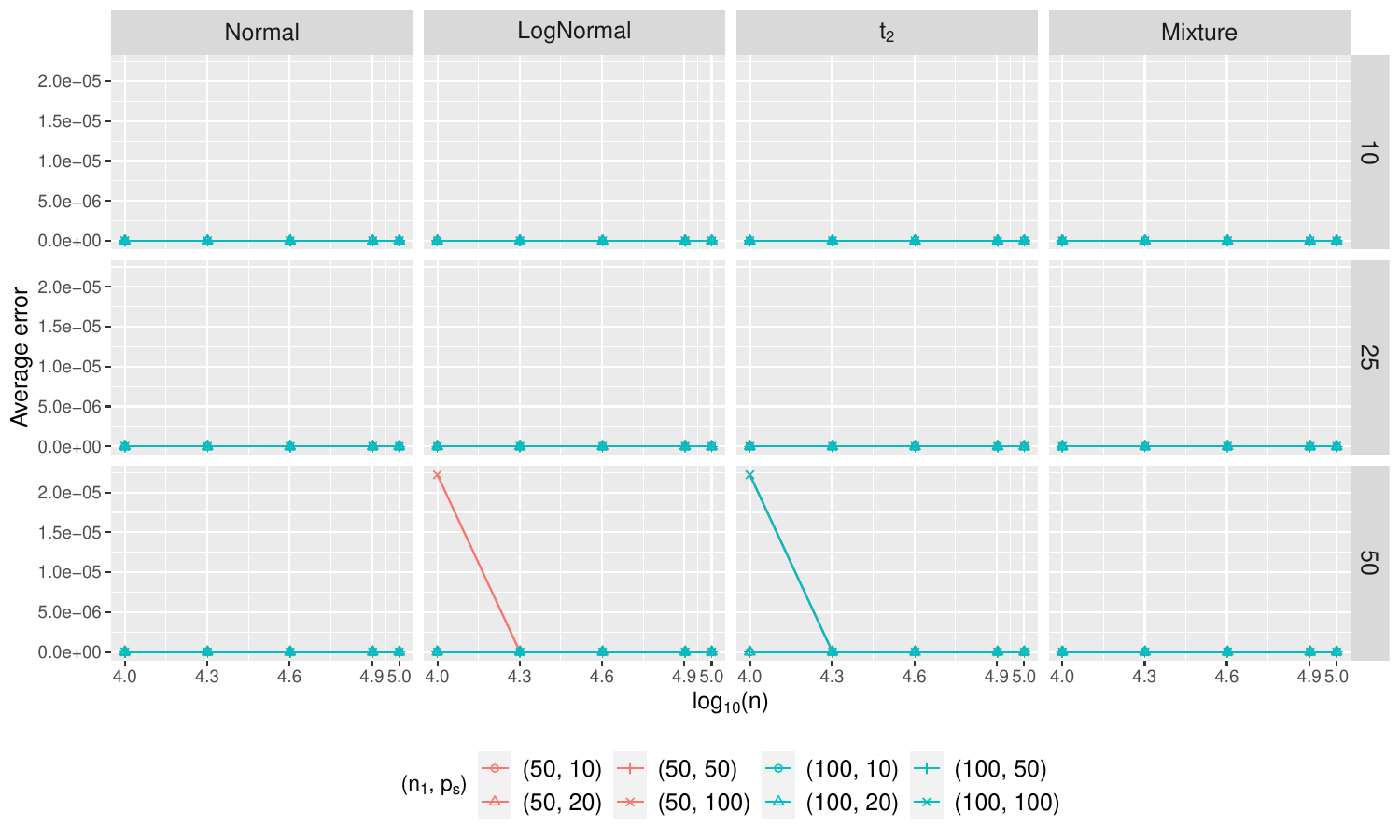} 
	\caption{Average error for $s = 1000$, $\boldsymbol{\Sigma} = \textbf{I}_p $ and $p_1 = 10, 25, 50$.}
	\label{errorI}
\end{figure}

\begin{figure}[!htb] 
	\centering 
	\includegraphics[width=1\textwidth]{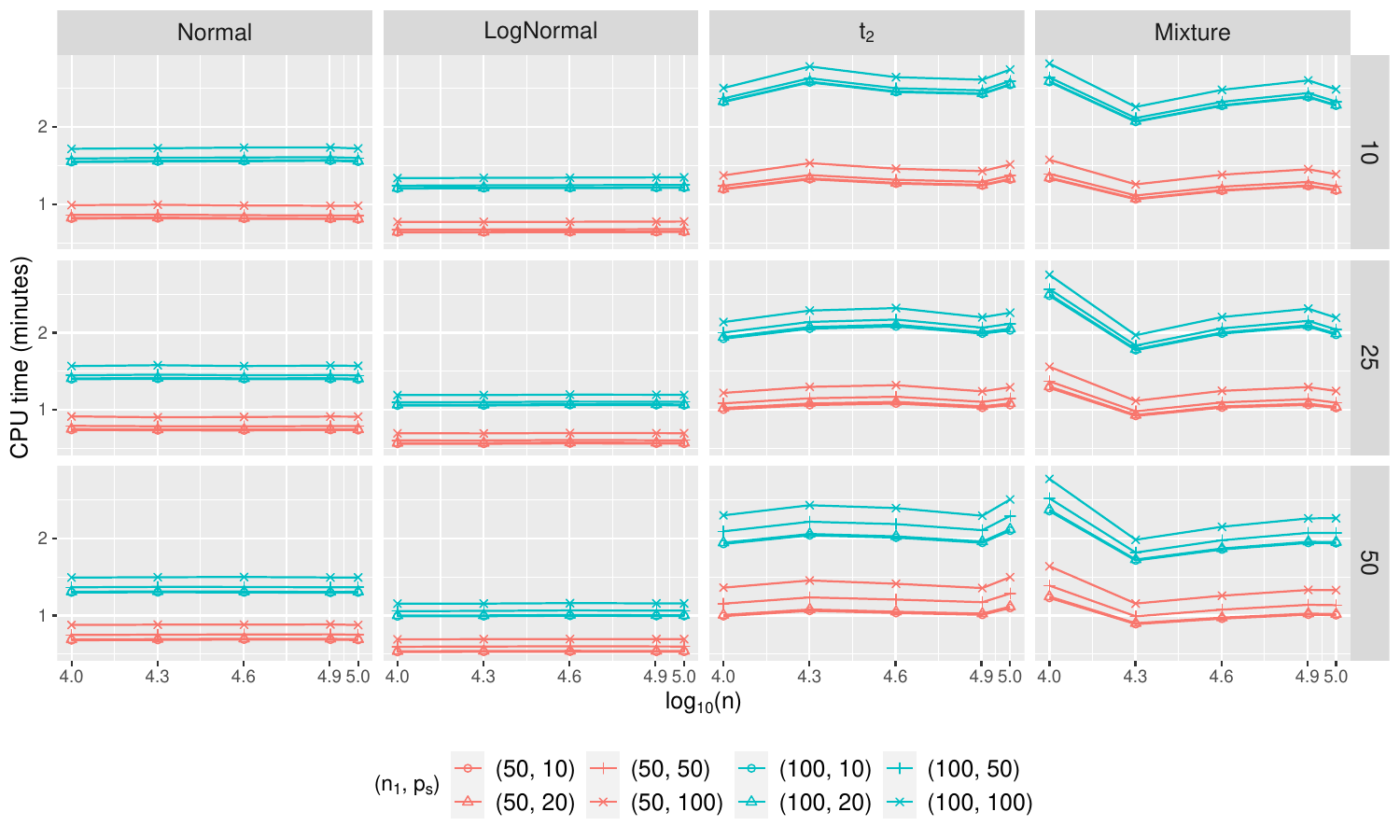} 
	\caption{CPU times (minutes) for $s = 1000$, $\boldsymbol{\Sigma} = \textbf{I}_p $ and $p_1 = 10, 25, 50$.}
	\label{timeI}
\end{figure}

\begin{figure}[!htb] 
	\centering 
	\includegraphics[width=1\textwidth]{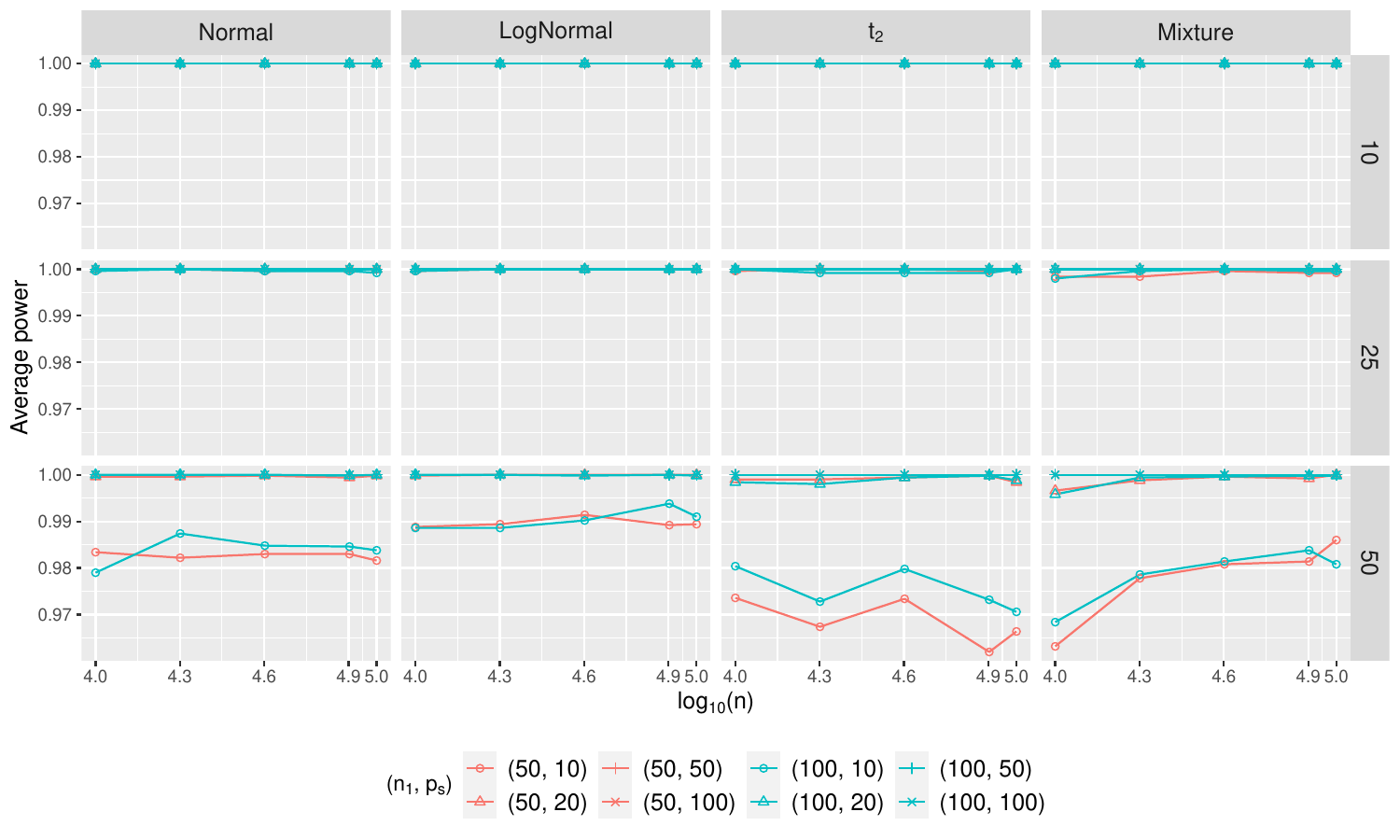} 
	\caption{Average power for $s = 1000$, $\boldsymbol{\Sigma} = 0.5\left(\textbf{J}_p+\textbf{I}_p\right) $ and $p_1 = 10, 25, 50$.}
	\label{power05}
\end{figure}

\begin{figure}[!htb] 
	\centering 
	\includegraphics[width=1\textwidth]{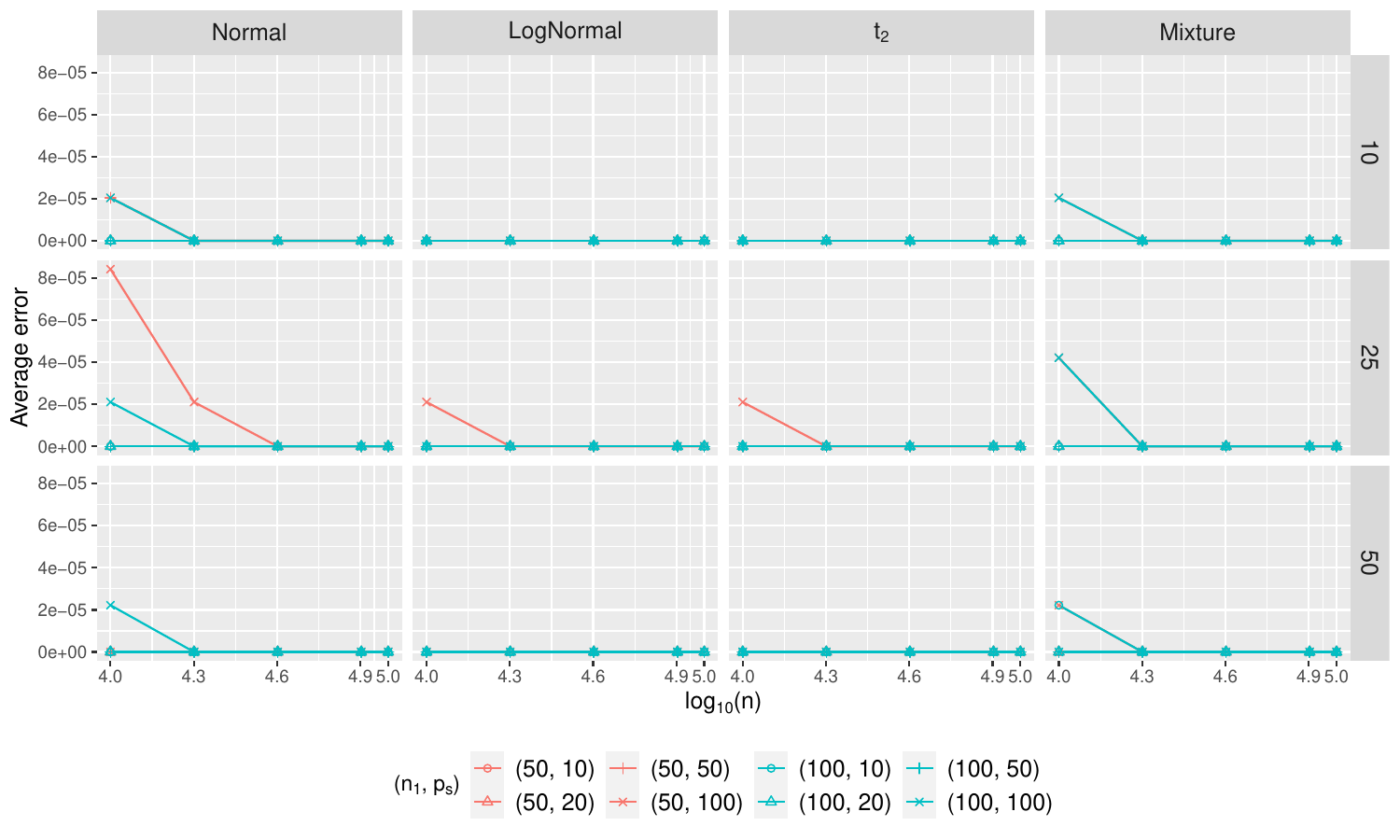} 
	\caption{Average error for $s = 1000$, $\boldsymbol{\Sigma} = 0.5\left(\textbf{J}_p+\textbf{I}_p\right) $ and $p_1 = 10, 25, 50$.}
	\label{error05}
\end{figure}

\begin{figure}[!htb] 
	\centering 
	\includegraphics[width=1\textwidth]{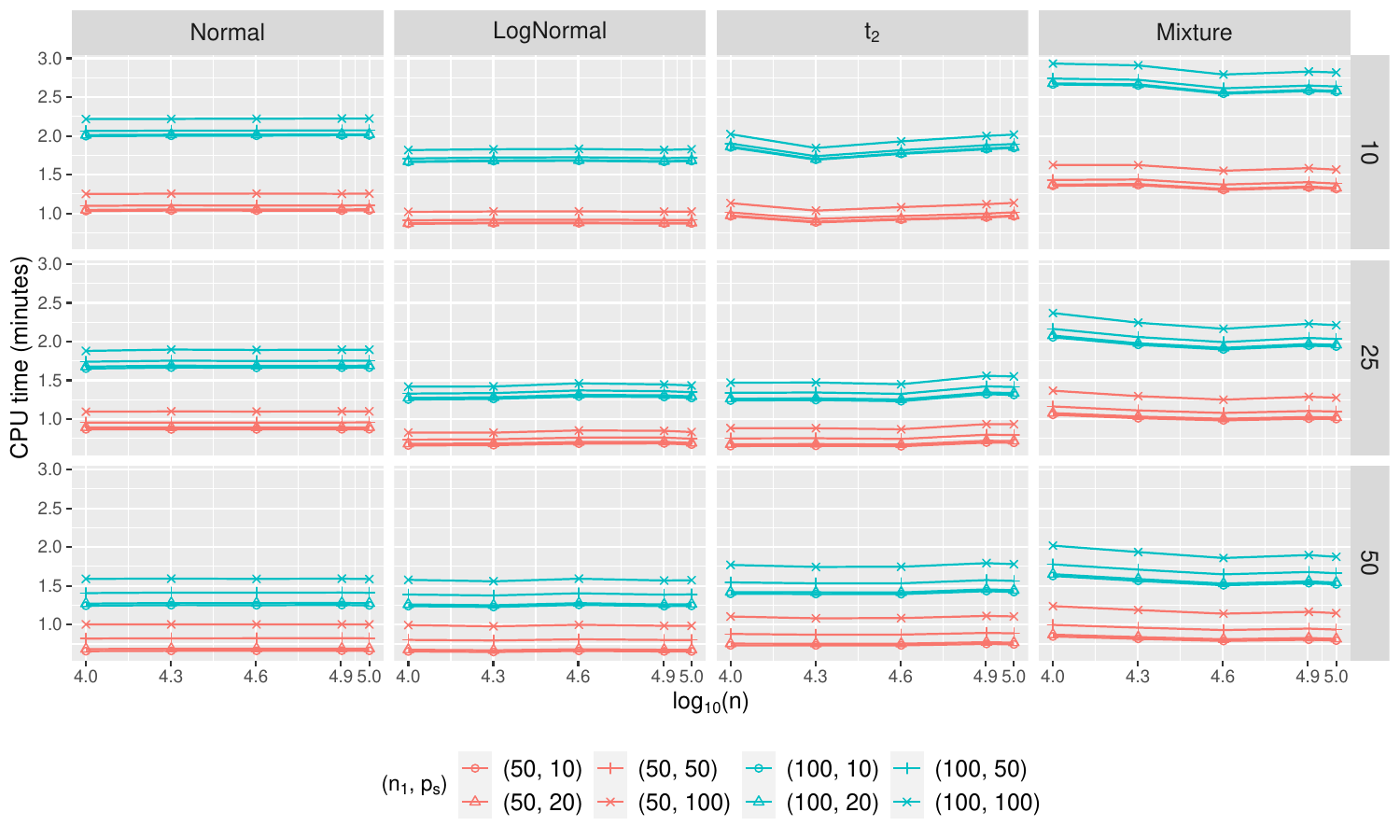} 
	\caption{CPU times (minutes) for $s = 1000$, $\boldsymbol{\Sigma} = 0.5\left(\textbf{J}_p+\textbf{I}_p\right) $ and $p_1 = 10, 25, 50$.}
	\label{time05}
\end{figure}

We use $s=1000$ as in \cite{singh}, focusing on $n_1$, $n_2$ and $p_s$. In Figures \ref{powerI} and \ref{power05} we provide the average power, in Figures \ref{errorI} and \ref{error05} the average error, and in Figures \ref{timeI} and \ref{time05} the computing time for the simulation experiment described in Section 4. We have that $n_1=50, 100$, $n_2=100$, and $p_s=10,20,50,100$.

Based on the results, we recommend to apply Algorithm 1 using $n_1=50$, $n_2=100$ and $p_s=0.1p$. When the value of $p_s$ is equal to $50$ and $100$, the increase of $n_1$ from $50$ to $100$ leads to increase of computing time while the average power and the average error are equal to one and zero,  respectively, for the most of the cases. Also, for both $n_1=50$ and $n_1=100$, the average error is not always equal to zero when $p_s=100$. Moreover, when the value of $p_s$ is equal to $10$ and $20$, the average power is less than one and the average error is more than zero for both $n_1=50$ and $n_1=100$. 

\newpage

\newpage

\newpage

\end{document}